\documentclass[11pt]{article}
\usepackage{epsfig}
\usepackage{verbatim}
\usepackage{amsfonts}
\usepackage{amsmath}
\usepackage{amssymb}
\usepackage{amsthm}
\usepackage{newlfont}
\usepackage{natbib}
\usepackage{setspace}
\usepackage{longtable}
\usepackage{color,xcolor}
\usepackage{authblk}
\usepackage [latin1]{inputenc}

\makeatletter 
\def\singlespace{\def\baselinestretch{1}\@normalsize}

\theoremstyle{definition}

\begin{document}
\title{These Unprecedented Times: The Dynamic Pattern Of COVID-19 Deaths Around The World}
\author{Zixuan Han}
\author{Tao Li\thanks{Corresponding author. Email: li.tao@mail.shufe.edu.cn}}
\author{Jinghong You}
\affil{\footnotesize School of Statistics and management, Shanghai
University of Finance and Economics, Shanghai, China}
%\author{}
\date{}

\maketitle
%%%%%%%%%%%%%%%%%%%%%%%%%%%%%%%%%%%%%%%%%%%%%%%%%%%%%%%%%%%%%%%%%%%%%%%%%%%%%%%%%%%%%%%%%%%%%%%%%%%%%%%%%%%%%%%%%%
\begin{abstract}
In this article, we deal with COVID-19 data to study the trend of the epidemic at the global situation. Choosing the mortality rate as an appropriate metric which measures the relative relation between the cumulative confirmed cases and death cases, we utilize the modified kernel estimator for random density function \citep{pm16} to obtain the density of the mortality rate, and apply Fr\'echet change point detection \citep{dm20} to check if there is any significant change on the dynamic pattern of COVID-19 deaths. The analysis shows that the pattern of global COVID-19 mortality rate seems to have an obvious change at 104 days after first day when the reported death cases exceed 30. Then we discuss the reasons of abrupt change of mortality rate trend from aspects both subjective and objective.
\end{abstract}

%%%%%%%%%%%%%%%%%%%%%%%%%%%%%%%%%%%%%%%%%%%%%%%%%%%%%%%%%%%%%%%%%%%%%%%%%%%%%%%%%%%%%%%%%%%%%%%%%%%%%%%%%%%%%%%%%%
\section{Introduction}

On February 11, 2020, the disease caused by novel corona virus was officially named as `COVID-19'. With the increase in the infection range and transmission speed, the World Health Organization declared the COVID-19 a pandemic on March 11, 2020. The number of confirmed cases and deaths increased sharply, which leads to the deterioration beyond expectations. By now, more than 40 millions cases were reported worldwide, leading to about 1113750 deaths in almost all countries.

The COVID-19 data have been studied a lot from the perspective of either epidemically or statistics. \cite{h20} applied the generalized logistic distribution with the SIR model to propose a more accurate and stable method for global infectious disease prediction. To quantify the confirmed cases and deaths longitudinally, \cite{cbc20}, \cite{r20} and \cite{ms20} modeled the cumulative trajectories across countries, while \cite{rh20} utilized a Monte Carlo approach to present the time dynamics. Furthermore, \cite{rk20} proposed time-varying auto-regressive models to analyze the count time series of daily new cases. Since the COVID-19 data in the United States present obvious trend in growth, \cite{zl20} proposed a method based on the generalized k-means to group the state-level time series patterns for the outbreak in the country, \cite{jm20} introduced a flexible algorithm using trajectories to determine the second surge behavior of epidemic in each state, and \cite{twz20} used the functional principal component analysis to study the data in the United States and tried to make further forecasts.

As time goes by, the trends of global surge in both confirmed and death cases of the epidemic in the period seem to have an obvious change, which naturally motivates us to find a measure combining the information from both cases to present a relative change trend. It is important for WHO to know well changes in the trend and distributions of global COVID-19 related mortality rate. In this article, we study the dynamic pattern of COVID-19 by treating the daily country-level data as the observations of random density functions, and utilize Fr\'echet change-point detection \citep{dm20}  to infer the presence of change point of dynamic pattern of worldwide mortality rate density, and analyze the reason from both objective  and subjective perspectives.

The rest of the article is organized as follows. In Section 2, we introduce the collection of COVID-19 data and make a clear description of the variables of interests. In Section 3, with the modified kernel type density estimator, we present the dynamic pattern of density function of mortality rate. In Section 4, with the methodology of change point detection for data of density functions, we present the change point of density functions of mortality rate, and try to analyze the reason to this phenomenon. A discussion follows in Section 5.
%%%%%%%%%%%%%%%%%%%%%%%%%%%%%%%%%%%%%%%%%%%%%%%%%%%%%%%%%%%%%%%%%%%%%%%%%%%%%%%%%%%%%%%%%%%%%%%%%%%%%%%%%%%%%%%%%%
%%%%%%%%%%%%%%%%%%%%%%%%%%%%%%%%%%%%%%%%%%%%%%%%%%%%%%%%%%%%%%%%%%%%%%%%%%%%%%%%%%%%%%%%%%%%%%%%%%%%%%%%%%%%%%%%%%
\section{Data collection and description}

Johns Hopkins University provides COVID-19 data worldwide, with a tracking map to present the global trend as well, which are publicly available at \textbf{https://www.jhu.edu/}. The raw data contains the daily confirmed and death cases for each country or region. In what follows, the daily cumulative confirmed and death cases can be obtained effortlessly. In this article, we focus on the mortality rate, which is defined as the number of daily cumulative death cases divided by the number of daily cumulative confirmed cases. The data set we have is the mortality rate of 189 countries for a period of 150 days from January 22, 2020 to September 1, 2020, each after the first date the country reports at least 30 death cases, which is defined as the origin of the time scale.

\begin{figure}[!ht]\centering
	\includegraphics[width=12cm]{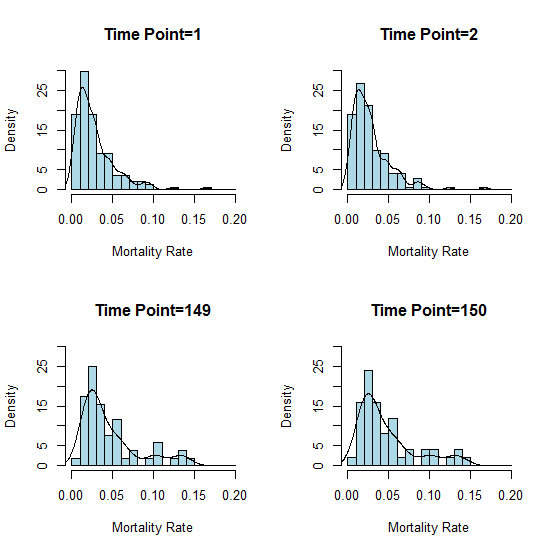}
	\caption{The histograms of mortality rate in the first (Time Point=1,2) and last two days (Time Point=149,150) of COVID-19 data.}\label{fg:histograms}
\end{figure}

As an example, the histograms of mortality rate in 189 countries on the first and last two days are presented in Figure \ref{fg:histograms}. From the figure we can see that all histograms are right-skewed. The densities on the first two days are quite similar and have a smaller mode than those on the last two days

Next, we utilize the modified kernel method \citep{pm16} to estimate the density of the mortality rate, and Fr\'echet change point detection to check if there is any significant change on the pattern of COVID-19 mortality rate.

%%%%%%%%%%%%%%%%%%%%%%%%%%%%%%%%%%%%%%%%%%%%%%%%%%%%%%%%%%%%%%%%%%%%%%%%%%%%%%%%%%%%%%%%%%%%%%%%%%%%%%%%%%%%%%%%%%
%%%%%%%%%%%%%%%%%%%%%%%%%%%%%%%%%%%%%%%%%%%%%%%%%%%%%%%%%%%%%%%%%%%%%%%%%%%%%%%%%%%%%%%%%%%%%%%%%%%%%%%%%%%%%%%%%%
\section{Dynamic pattern of density of mortality rate}

\subsection{Methodology}

Generally speaking, we assume that data consists of $ T $ random density functions $ f_{1}(x),\cdots,f_{T}(x) $, which have common support $ S=[\alpha_{1},\alpha_{2}] $. However, in practice, we may just observe an $ i.i.d. $ sample instead of densities directly, which consists of data $ W_{t1},\cdots,W_{tn_{t}} $, generated from the random density $ f_{t}(x)$, $t=1,\cdots,T $. Under this circumstance, the information about each specific density can be acknowledged only through the observed random sample, therefore, these densities should be estimated at first.

Since the densities are supported on a compact intervals, the performance of standard kernel density estimators at the boundary may not be as well as expected. To rectify the boundary effects, \cite{pm16} proposed a modified density estimator of kernel type which satisfies the consistency properties in the space of density functions.

Let $ f(x) $ be a continuous probability density function supported on $ S $, without loss of generality, we take $ S= $[0,1], and $ \mathcal{K} $ is the corresponding kernel with bandwidth $ h<1/2 $. Here, the kernel $ \mathcal{K} $ with bounded variation is symmetric of 0, satisfying the conditions that $ \int_{0}^{1}\mathcal{K}(u)du>0 $, $ \int_{\mathcal{R}}^{}|u|\mathcal{K}(u)du $, $ \int_{\mathcal{R}}^{}\mathcal{K}^{2}(u)du $ and $ \int_{\mathcal{R}}^{}|u|\mathcal{K}^{2}(u)du $ are finite. Let $ \mathcal{F} $ be the class of univariate probability density function $ f $, satisfying  $\int_{\mathcal{R}}^{}u^{2}f(u)du<\infty $. Then for $ x\in[0,1] $, the modified kernel estimator of $ f\in \mathcal{F} $ from an $i.i.d.$ sample $ W_{1},\cdots,W_{n}\sim f $ is defined as
\begin{equation}
\hat f(x)=\sum_{l=1}^{n}\mathcal{K}\left(\dfrac{x-W_{l}}{h}\right)w(x,h)\biggl/\sum_{l=1}^{n}\int_{0}^{1}\mathcal{K}\left(\dfrac{y-W_{l}}{h}\right)w(y,h)dy,
\end{equation}
where the weight function $ w(x,h) $ is defined as follows:
\begin{equation}
w(x,h)=\begin{cases}
\left(\int_{-x/h}^{1}\mathcal{K}(u)du\right)^{-1}, & \text{for $ x\in[0,h), $}\\
\left(\int_{-1}^{(1-x)/h}\mathcal{K}(u)du\right)^{-1}, & \text{for $ x\in(1-h,1], $}\\
1, & \text{otherwise}.
\end{cases}
\end{equation}

%%%%%%%%%%%%%%%%%%%%%%%%%%%%%%%%%%%%%%%%%%%%%%%%%%%%%%%%%%%%%%%%%%%%%%%%%%%%%%%%%%%%%%%%%%%%%%%%%%%%%%%%%%%%%%%%%%
%%%%%%%%%%%%%%%%%%%%%%%%%%%%%%%%%%%%%%%%%%%%%%%%%%%%%%%%%%%%%%%%%%%%%%%%%%%%%%%%%%%%%%%%%%%%%%%%%%%%%%%%%%%%%%%%%%
\subsection{Dynamic pattern of density of mortality rate}

For each time point $ t $, $ t=1, \cdots,150 $, the data set consists of $ n=189 $ country-level data as the observations from density function $ f_{t}  $ of mortality rate.

\begin{figure}[!ht]\centering
	\includegraphics[width=0.8\textwidth]{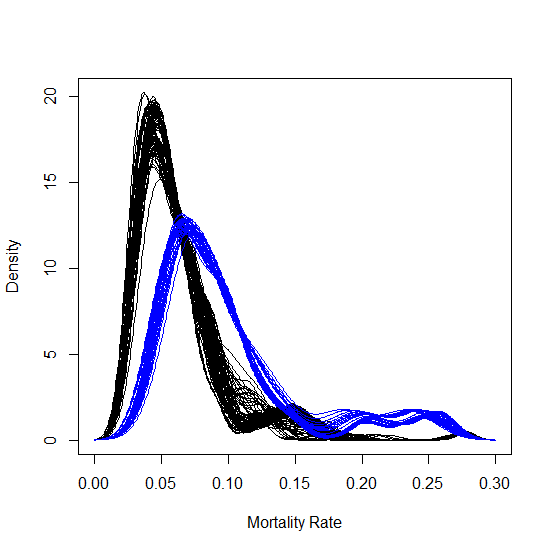}
	\caption{Daily mortality rate distributions represented as density functions during the period from time point 1 to 150 for the COVID-19 data.}\label{fg:density2D}
\end{figure}

\begin{figure}[!ht]\centering
	\includegraphics[width=0.8\textwidth]{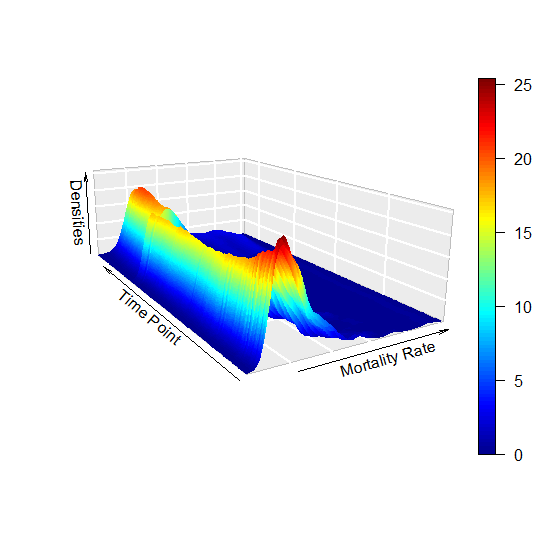}
	\caption{Time-dynamic daily mortality rate distributions represented as density functions during the period from time point 1 to 150 for the COVID-19 data.}\label{fg:density3D}
\end{figure}

Figure \ref{fg:density2D} displays the density curves of mortality rate over the period of 150 days. From the figure, it is clear that there exists two different types of densities. To see if this abrupt change on distributions happen at a time point, Figure \ref{fg:density3D} displays the densities in a three-dimensional space, adding the dimension of time scale. We can see from Figure \ref{fg:density3D} that this change point does exist. Moreover, the distributions before the change point have a small mode with small variance, while the ones after the change point indicate that the mortality rates gradually tend to larger with large variability. At the first stage when the virus outbreaks, the performances of different countries are quite similar. But along with the different response and medical supplies to the pandemic, they demonstrates the different results, especially the proportion of countries with high mortality rate increases.

%%%%%%%%%%%%%%%%%%%%%%%%%%%%%%%%%%%%%%%%%%%%%%%%%%%%%%%%%%%%%%%%%%%%%%%%%%%%%%%%%%%%%%%%%%%%%%%%%%%%%%%%%%%%%%%%%%
%%%%%%%%%%%%%%%%%%%%%%%%%%%%%%%%%%%%%%%%%%%%%%%%%%%%%%%%%%%%%%%%%%%%%%%%%%%%%%%%%%%%%%%%%%%%%%%%%%%%%%%%%%%%%%%%%%
\section{Change point detection for dynamic pattern}

\subsection{Methodology}
Change point detection is increasingly becoming an important tool in many fields, identifying whether there exists an abrupt change in a data sequence and determining its location. A lot of literatures on change point detection for univariate and multivariate data have been studied.

Nowadays, data consisting of random variables which cannot be considered as univariate or multivariate data arise commonly, indicating that the former approaches may no longer applicable to these data. To address this problem, further improvements have been developed in this area. \cite{pc15}, \cite{dvr16} and \cite{wyr18} provided parametric approaches in networks, \cite{ty06} and \cite{kym07} for time series data,
\cite{ddd05}, \cite{ach12} and \cite{ga18} made detection using kernel density, \cite{ks12} with density-ratio estimation, \cite{mz03} utilized sigular-spectrum analysis, \cite{cz15}, \cite{cf17} and \cite{cc19} proposed graph-based approaches. However, all of the methods mentioned above own their limits and constraints to be applied to a general metric space. To solve this problem, \cite{dm20} introduced a method of change points detection for independent data sequence taking values in a general metric space.

The method proposed by \cite{dm20} is based on the main ideas of the two sample test for random objects introduced in \cite{cf17}, presenting the difference in Fr\'echet means or Fr\'echet variance \citep{pm19} for sample of data taking values in a general metric space. Specifically, it utilizes the Fr\'echet means and Fr\'echet variance before and after the change point to provide a test for the presence and obtain the estimation of locations.

Let $ Y_{1},\cdots,Y_{n} $ be a sequence data values in a metric space $ (\varOmega,\mathcal{D}) $, given $ k $, $0<k<1 $, suppose the location of the change point is at $ [nk] $, then we partition the data sequence into two segments, indicating $ Y_{1},\cdots,Y_{[nk]}\sim F_{1} $ and $ Y_{[nk]+1},\cdots,Y_{n}\sim F_{2} $. Denote $ \mu_{1} $ and $ \mu_{2} $ be the Fr\'echet mean of $ F_{1} $ and $ F_{2} $, respectively, defined as
\begin{equation}
\mu_{1}=\mathop{\arg\min}_{w\in\varOmega}E_{1}(d^{2}(Y,w)), Y\sim F_{1}, \quad
\mu_{2}=\mathop{\arg\min}_{w\in\varOmega}E_{2}(d^{2}(Y,w)), Y\sim F_{2},
\end{equation}
and the corresponding Fr\'echet variance are defined as
\begin{equation}
V_{1}=\mathop{\min}_{w\in\varOmega}E_{1}(d^{2}(Y,w)), Y\sim F_{1}, \quad
V_{2}=\mathop{\min}_{w\in\varOmega}E_{2}(d^{2}(Y,w)), Y\sim F_{2},
\end{equation}
where $ E_{1}(\cdot) $ and $ E_{2}(\cdot) $ are the expectations with respect to $ F_{1} $ and $ F_{2} $ respectively.

Then, the detection of change point is to determine whether the segments differ or not, meaningfully, whether $ \mu_{1}=\mu_{2}$ and $ V_{1}=V_{2} $, yielding the judgment of $ F_{1}=F_{2} $. Since computation of Fr\'echet mean and variance both need a minimum number of observations, then set $ k $ in $ I_{c}=[c,1-c] \subset [0,1] $ for some $ c>0 $. Therefore, the test of change point and the statistic can be designed and constructed based on the estimated Fr\'echet mean and variance. Detailed discussion is refer to \cite{dm20}.

In this article, we apply this method to data which consists of density functions. Under this circumstance, we use metric $ d_{w} $ to replace $ d $. First begin with a simple definition of Wasserstein metric $ d_{w}$. Let $ \mathcal{F} $ be the class of univariate probability density functions denoted before. For $ f,g\in \mathcal{F} $, $ F,G $ are the cumulative distribution functions of $ f $ and $ g $ respectively,  $ Q_{f} $ and $ Q_{g} $ are the corresponding quantile distribution functions. Then the Wasserstein-2 distance between these two distributions is defined as
\begin{equation}
d_{w}^{2}(f,g)=\int_{0}^{1}(Q_{f}(t)-Q_{g}(t))^{2}dt.
\end{equation}

Denote $ f_{1},\cdots,f_{n} $ be a data sequence of probability density functions in $ \mathcal{F} $, change point occurs at $ [nk] $, $ k\in I_{c} $, then
the estimated Fr\'echet mean and variance of the data before the change point are defined as
\begin{equation}
\hat{\mu_{1}}=\mathop{\arg\min}_{f\in \mathcal{F}}\frac{1}{[nk]}\sum_{i=1}^{[nk]}d_{w}^{2}(f_{i},f), \quad
\hat{V_{1}}=\frac{1}{[nk]}\sum_{i=1}^{[nk]}d_{w}^{2}(f_{i},\hat{\mu_{1}}),
\end{equation}
and similar computation for data after the change point, i.e.,
\begin{equation}
\hat{\mu_{2}}=\mathop{\arg\min}_{f\in \mathcal{F}}\frac{1}{n-[nk]}\sum_{i=[nk]+1}^{n}d_{w}^{2}(f_{i},f), \quad
\hat{V_{2}}=\frac{1}{n-[nk]}\sum_{i=[nk]+1}^{n}d_{w}^{2}(f_{i},\hat{\mu_{2}}).
\end{equation}

Next, define contaminated Fr\'echet variance(\cite{dm20}) by replacing the Fr\'echet mean with another one from complementary segment, namely,
\begin{equation}
\hat{V^{c}_{1}}=\frac{1}{[nk]}\sum_{i=1}^{[nk]}d_{w}^{2}(f_{i},\hat{\mu_{2}}), \quad
\hat{V^{c}_{2}}=\frac{1}{n-[nk]}\sum_{i=[nk]+1}^{n}d_{w}^{2}(f_{i},\hat{\mu_{1}}).
\end{equation}
Meanwhile, define
\begin{equation}
\hat{\sigma}^{2}=\frac{1}{n}\sum_{i=1}^{n}d_{w}^{4}(f_{i},\hat{\mu})-\hat{V},
\end{equation}
where
\begin{equation}
\hat{\mu}=\mathop{\arg\min}_{f\in \mathcal{F}}\frac{1}{n}\sum_{i=1}^{n}d_{w}^{2}(f_{i},f), \quad
\hat{V}=\frac{1}{n}\sum_{i=1}^{n}d_{w}^{2}(f_{i},\hat{\mu}).
\end{equation}

Then, for $ k\in I_{c} $, the proposed test statistic is constructed as
\begin{equation}
T_{n}(k)=\dfrac{k(1-k)}{\hat{\sigma}^{2}}\{(\hat{V_{1}}-\hat{V_{2}})^{2}+(\hat{V^{c}_{1}}-\hat{V_{1}}+\hat{V^{c}_{2}}-\hat{V_{2}})^{2}\},
\end{equation}
which is considered as a function of $ k $.

In the above equation, it is natural to consider $ \hat{V^{c}_{1}}-\hat{V_{1}} $ and $ \hat{V^{c}_{2}}-\hat{V_{2}} $ as the between group variance of two segments, thus, under the hypothesis that change point exists in the sequence,  $ \hat{V^{c}_{1}}-\hat{V_{1}} $ and $ \hat{V^{c}_{2}}-\hat{V_{2}} $ will no longer be small, and the absolute value of difference $ |\hat{V_{1}}-\hat{V_{2}}| $ is expected to be large. Therefore, the statistic $ T_{n}(k) $ incorporates both differences in Fr\'echet means and Fr\'echet variance of data segments for change point detection. Namely, change point is the location which maximizes the above differences. More detailed theoretical results are provided in \cite{dm20}.

%%%%%%%%%%%%%%%%%%%%%%%%%%%%%%%%%%%%%%%%%%%%%%%%%%%%%%%%%%%%%%%%%%%%%%%%%%%%%%%%%%%%%%%%%%%%%%%%%%%%%%%%%%%%%%%%%%
%%%%%%%%%%%%%%%%%%%%%%%%%%%%%%%%%%%%%%%%%%%%%%%%%%%%%%%%%%%%%%%%%%%%%%%%%%%%%%%%%%%%%%%%%%%%%%%%%%%%%%%%%%%%%%%%%%
\subsection{Change point detection for dynamic pattern}

\begin{figure}[!ht]\centering
	\includegraphics[width=0.8\textwidth]{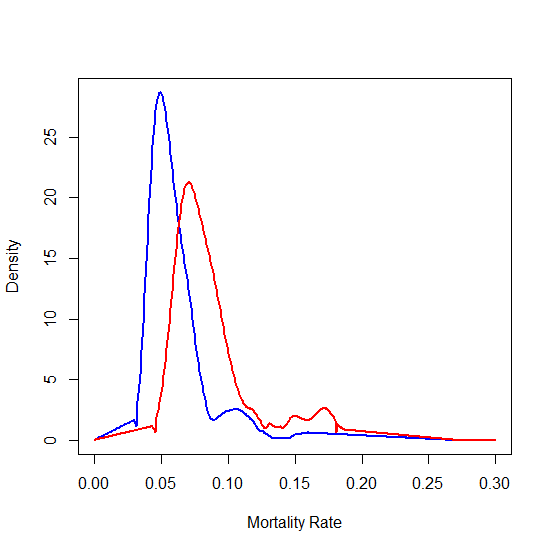}
	\caption{Estimated Fr\'echet mean densities of mortality rate before (blue) and after (red) the estimated change point 104 for COVID-19 death data.}\label{fg:Frechet mean}
\end{figure}

For the mortality rate data, we apply the change point detection method based on the Fr\'echet variance of density functions, with time point be the index for the sequence of distributions. Figure \ref{fg:T_{n}(u)} presents the result of test statistic $ T_{n}(k) $, indicating that there indeed exists a change point in the data, which maximizes the differences between segments before and after this location. The estimated location of change point is time point 104. To have a comprehensive understanding, Figure \ref{fg:Frechet mean} shows the estimated Fr\'echet means of distributions before and after time point 104. The clear differences between two curves in both location and scale from the figure also serve as the evidence of the presence of change point.

\begin{figure}[!ht]\centering
	\includegraphics[width=0.8\textwidth]{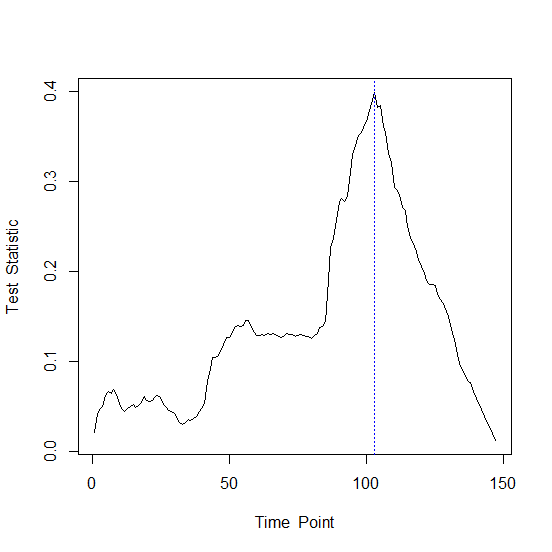}
	\caption{The function of test statistic $ T_{n}(k) $ . The dotted line indicates the estimated location of change point at point 104.}\label{fg:T_{n}(u)}
\end{figure}

Since we have detected the presence of change point and determine the location, it is natural to figure out the possible factors that may affect and cause the abrupt change. In fact, time point 104 is about three and a half months after the date when death cases exceeded 30 for the first time in each country. During the first three and a half months, the mortality rates clustered at smaller values due to various reasons. For one thing, many countries and governments paid no or little attention on the infectious capacity and transmission speed of the virus, and too confident about their own public health systems and medical technology, combining with the expensive cost for test which should be undertaken by patients themselves, leading to the inadequate number of tests seriously. On the other hand, owing to the limited level of medical service conditions, some countries and regions were unable to provide enough tests for corona virus timely, yielding the small screening scale for potential patients. The reasons discussed above may all contribute to the epidemic situation, however, it seems that some unrevealed factors may exist for spread and infection of virus in large scale in a short time, leading to the outbreak of corona virus all over the world.

\begin{figure}[!ht]\centering
	\includegraphics[width=\textwidth]{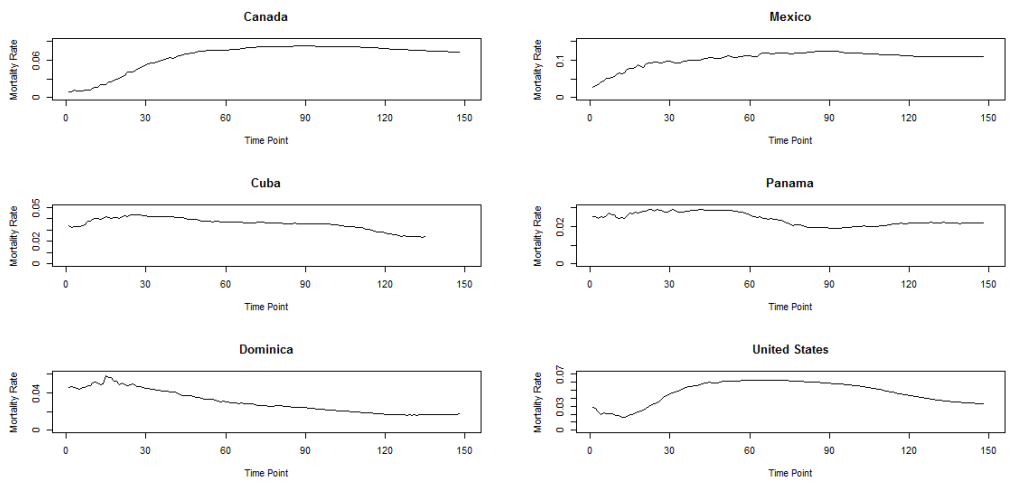}
	\caption{Time series of mortality rate in several North American countries.} \label{fg:NorAm}
\end{figure}

\begin{figure}[!ht]\centering
	\includegraphics[width=\textwidth]{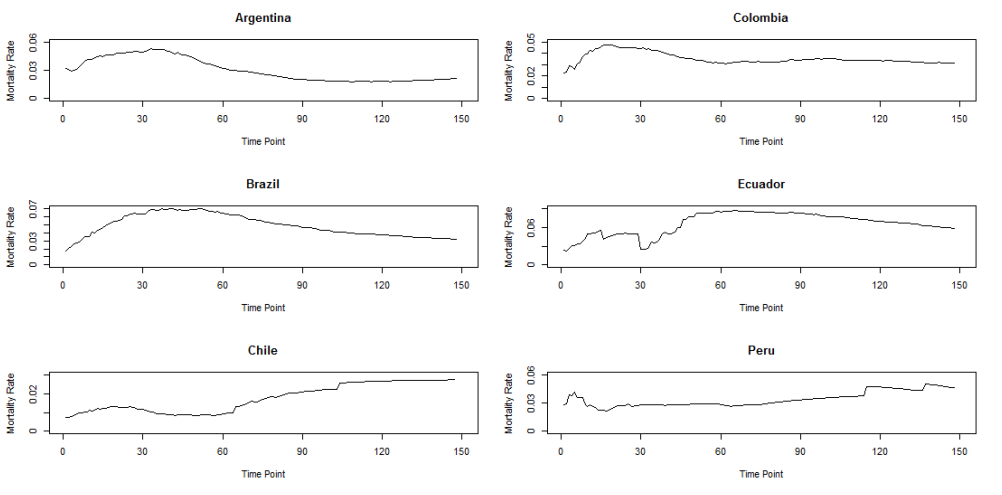}
	\caption{Time series of mortality rate in several South American countries.} \label{fg:SouAm}
\end{figure}

\begin{figure}[!ht]\centering
	\includegraphics[width=\textwidth]{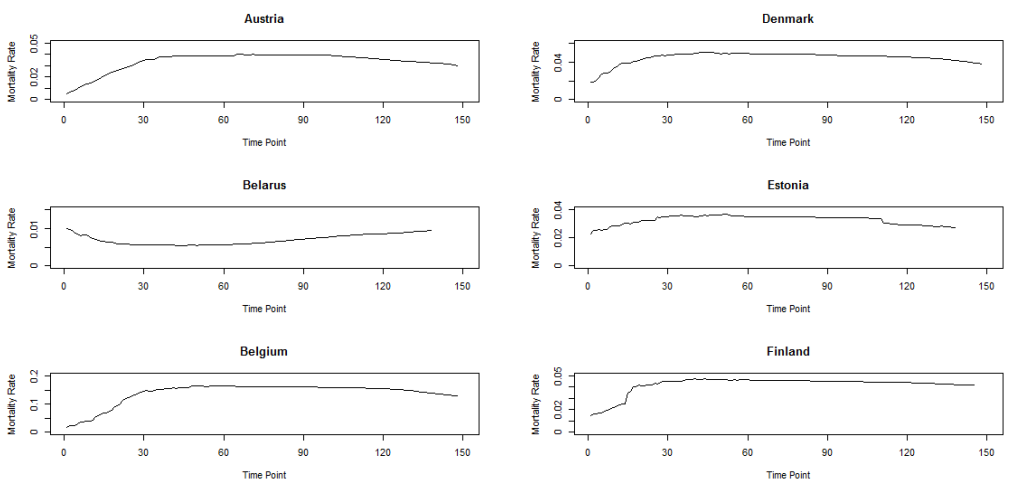}
	\\
	\includegraphics[width=\textwidth]{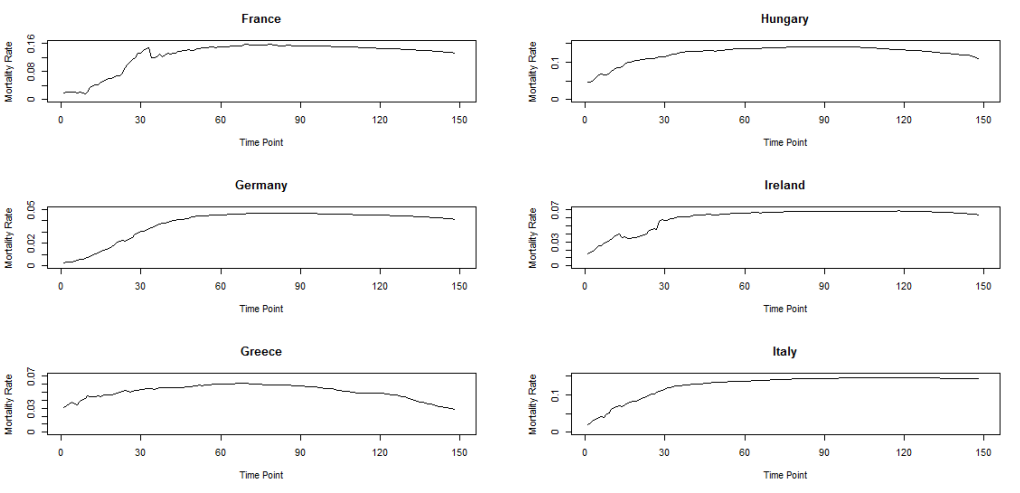}
	\caption{Time series of mortality rate in several European countries.} \label{fg:Europe1}
\end{figure}

\begin{figure}[!ht]\centering
	\includegraphics[width=\textwidth]{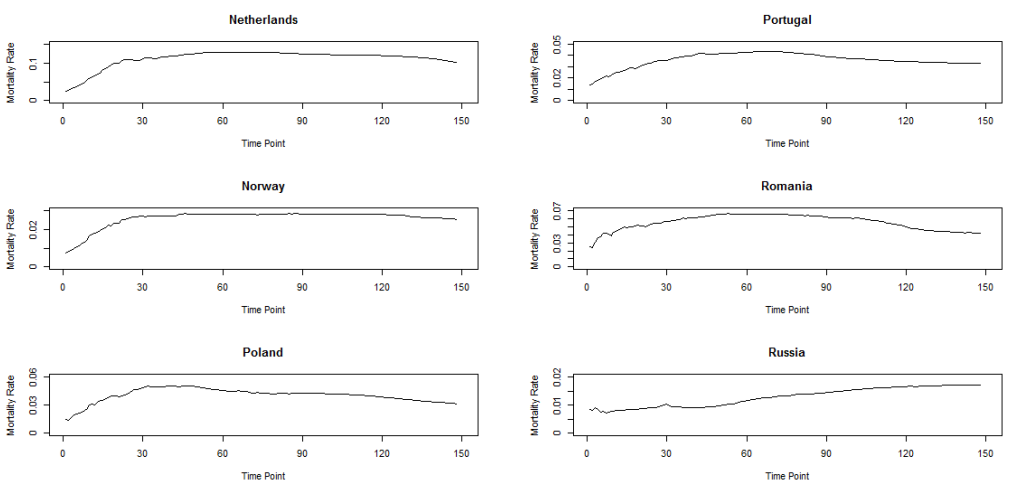}
	\\
	\includegraphics[width=\textwidth]{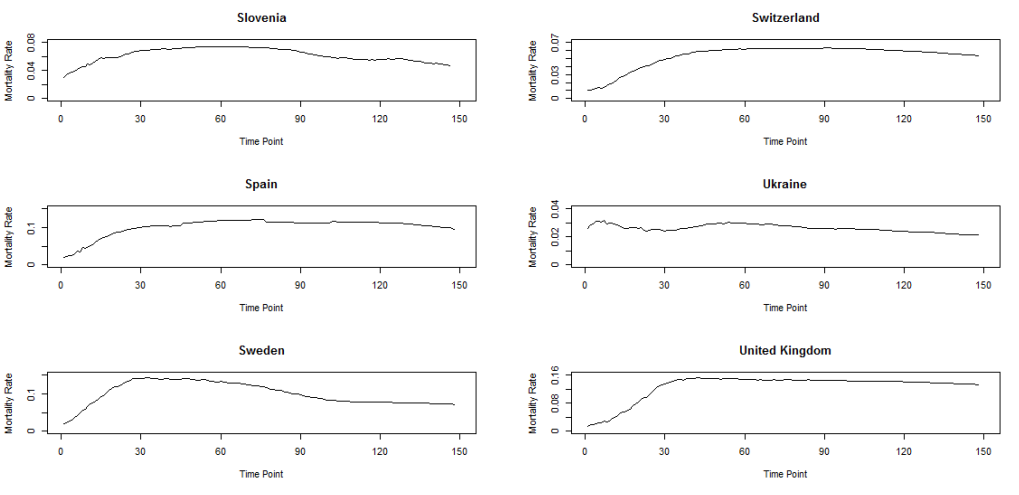}
	\caption{Time series of mortality rate in several European countries(continued).} \label{fg:Europe2}
\end{figure}

\begin{figure}[!ht]\centering
	\includegraphics[width=\textwidth]{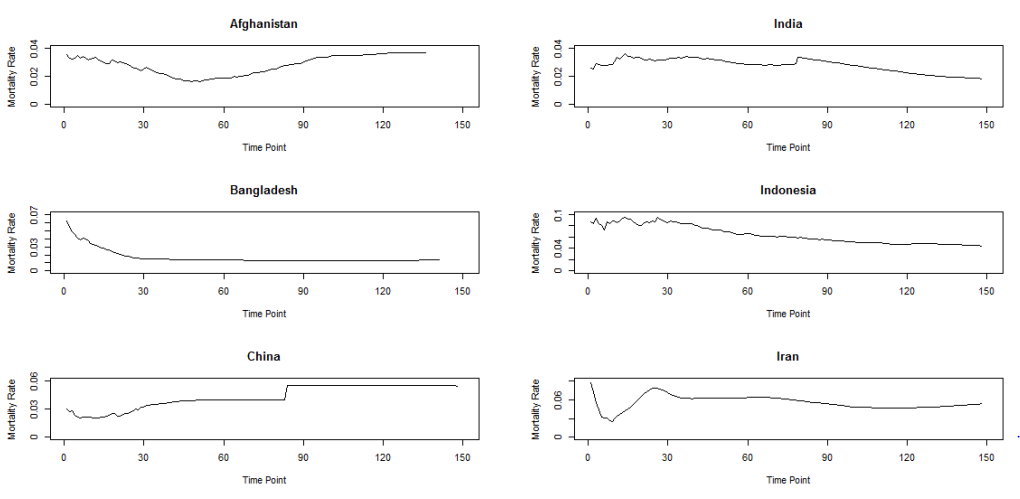}
	\\
	\includegraphics[width=\textwidth]{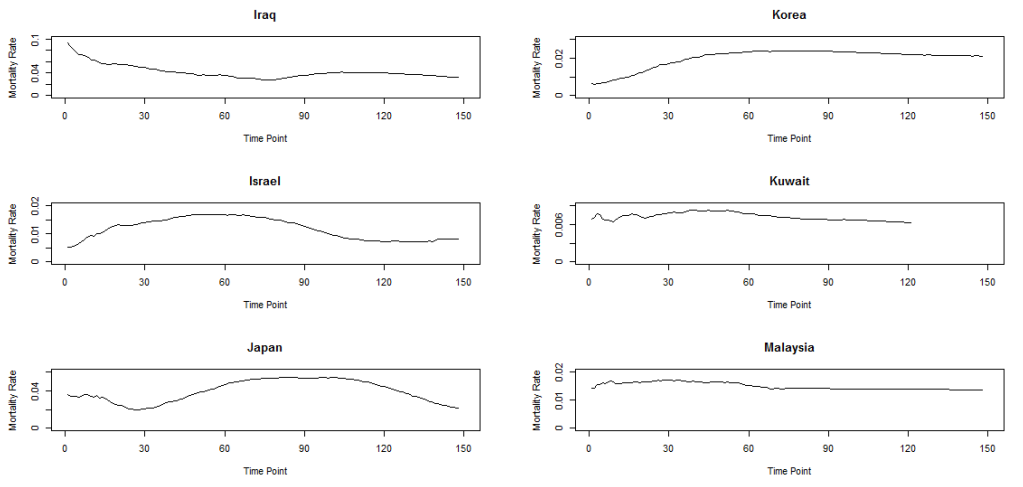}
	\caption{Time series of mortality rate in several Asian countries.} \label{fg:Asia}
\end{figure}

\begin{figure}[!ht]\centering
	\includegraphics[width=\textwidth]{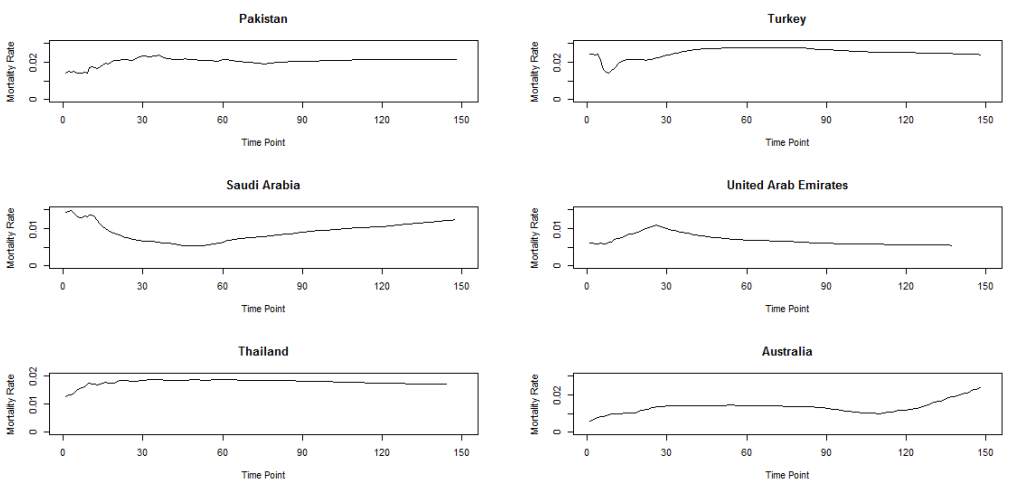}
	\caption{Time series of mortality rate in several Asian and Oceania countries.} \label{fg:AsOcea}
\end{figure}

\begin{figure}[!ht]\centering
	\includegraphics[width=\textwidth]{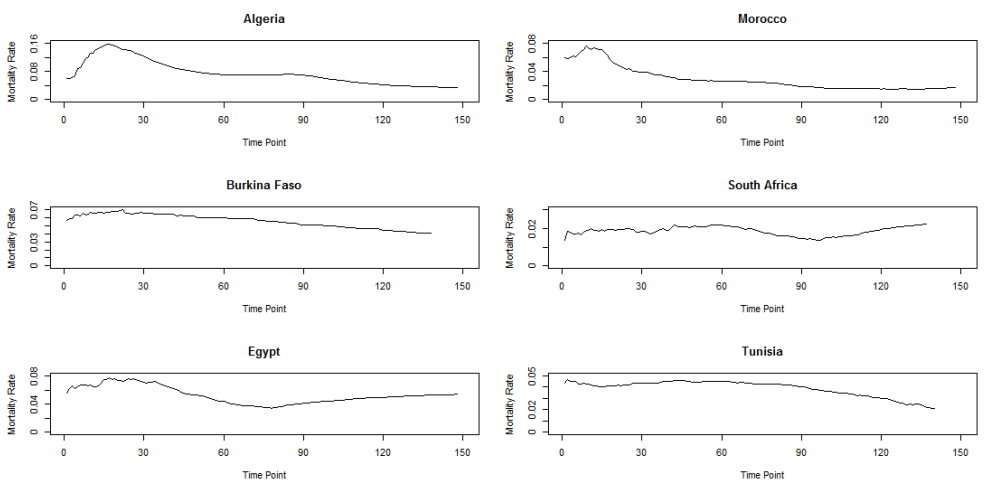}
	\caption{Time series of mortality rate in several African countries.} \label{fg:Africa}
\end{figure}

\begin{figure}[!ht]\centering
	\includegraphics[width=0.8\textwidth]{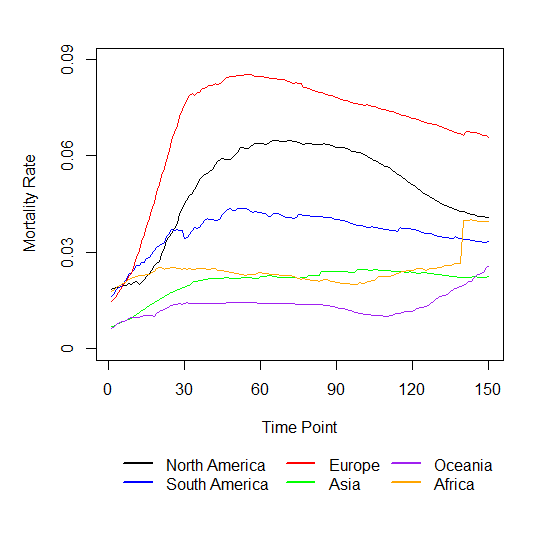}
	\caption{Time series of mortality rate in different continents.} \label{fg:avgall}
\end{figure}

To show the change of time trend of mortality rate before and after 104 days in detail, we select some countries from different continents and draw the time series of mortality rate in each selected country in Figures \ref{fg:NorAm}-\ref{fg:Africa}. We can see that at the first stage, the mortality rate curves in most countries have the pattern of climbing up; some countries climb up first and then get stable; and few countries has the pattern of declining. But at the second stage, the performances are quite different among all selected countries. Some countries are still climbing such as Australia, Saudi Arabia, South Africa and so on; some countries get little change such as China, Italy and so on; some countries have a pattern of declining such as Dominica, United State, Japan and so on. Besides, we can that some European countries have surprising high mortality rate such as France, United Kingdom, Hungary and Italy. To make an overall evaluation on the performance of mortality rate among different continents, Figure \ref{fg:avgall} depicts the time series of overall mortality rate in each continent. It is clear that the mortality in Europe is much higher than the others.

To further study the dynamic pattern of COVID-19 deaths, with the estimation of the density functions of mortality rate, we calculate the proportion of different continents in all countries whose mortality rate is smaller than 5 or 10 percentile or larger than 90 or 95 percentile of the mortality rate distribution for each day, and take the average of the proportion for the first 104 days and the last 46 days. The result is presented by Table \ref{tab:region}.  From the table and Figure \ref{fg:avgall}, we can see that the countries with high mortality rate ($> 90$ percentile or 95 percentile) mostly come from Europe no matter at the first stage (first 104 days) or the second stage (last 46 days). Moreover, the proportion of Europe in high mortality rate increases rapidly from the first stage to the second stage. Obviously, compared with the other countries, the emergency measure taken by European countries to the virus is not good enough to contain the disease. The proportion of North America in high mortality also has a slightly increase from the first stage to the second stage. At the first stage, Asia and Africa also account for the big proportion of high mortality rate. But different from the European, their proportions at the second stage decrease rapidly, especially the proportion of Asia at the second stage is near zero. Such big reduce of proportions may attribute to the powerful prevention measures and effective treatments discussed before. On the other hand, for the significant increase in Europe and North America, a research report announced by Brazil's Oswaldo Cruz Foundation in June may provide some explanations. In the report, the genetic studies of the new corona virus were compared with genetic map of new corona virus from all over the world preserved in the Global Initiative of Sharing All Influenza Data(GISAID), indicating that the new corona virus prevalent in Brazil was similar to samples from Europe, North America and Oceania. Later on July 2, a global research combining the achievements of several research institutions in the United States and the United Kingdom was conducted, according to CNN reports. The team collected corona virus samples from clinical patients in Europe and the United States and sequenced the genome, stating that the virus strain of type D first appeared in Europe and America had gene mutation. The infectivity of new mutant virus, which was named virus strain of type G, was three to nine times higher than that of the original one due to faster reproduction in the upper respiratory tract as note and throat, and had completely replaced the primary strain, becoming the main cause of disease in Europe and America at present. Moreover, some European countries have relaxed management after the prevention measures have achieved initial results, leading to signs of a rebound in epidemic situations. To avoid this phenomenon, governments called for strong control measures again, which triggered the massive demonstrations against the pandemic, providing opportunities for the rapid spread of the virus among the dense crowd. Meanwhile, the lack of medical supplies,  the overrun hospitals and the ineffective measures of epidemic prevention in some countries also further aggravated the situations. Therefore, all these complicated aspects discussed above formed the abrupt change in the global mortality rate.

\begin{table}[h]
	\centering
\caption{Averaged proportion of several continents at different quantiles of mortality rate before and after time point 78 (the number in the parenthesis is the number of countries included in the data set )} \label{tab:region}
	\begin{tabular}{ccccccccc}
		\hline 	
		& \multicolumn{2}{c}{5\%} & \multicolumn{2}{c}{10\%} & \multicolumn{2}{c}{90\%} & \multicolumn{2}{c}{95\%}\\\hline
		$ Continent $ &  $before$ & $after$ & $before$ & $after$ & $before$ & $after$ & $before$ & $after$ \\\hline			
		North America(17)  &	0.0796& 	0.0943& 	0.0705& 	0.0890& 	0.0973& 	0.1022& 	0.0839& 	0.1103\\
		South America(12)  &	0.1246& 	0.0766& 	0.1263& 	0.0731& 	0.0987& 	0.0534& 	0.1023& 	0.0479\\
		Europe(51)       &	0.0642& 	0.1409& 	0.0654& 	0.1468& 	0.4376& 	0.7935& 	0.4449& 	0.8162\\
		Asia(42)         &	0.1886& 	0.2476& 	0.1918& 	0.2709& 	0.1294& 	0.0023& 	0.1220& 	0.0000\\
		Oceania(18)      &	0.0182& 	0.0033& 	0.0196& 	0.0094& 	0.0000& 	0.0000& 	0.0000& 	0.0000\\
		Africa(49)       &	0.5248& 	0.4373& 	0.5264& 	0.4108& 	0.2370& 	0.0486& 	0.2468& 	0.0256\\\hline 	
	\end{tabular}
	
\end{table}	
%%%%%%%%%%%%%%%%%%%%%%%%%%%%%%%%%%%%%%%%%%%%%%%%%%%%%%%%%%%%%%%%%%%%%%%%%%%%%%%%%%%%%%%%%%%%%%%%%%%%%%%%%%%%%%%%%%
%%%%%%%%%%%%%%%%%%%%%%%%%%%%%%%%%%%%%%%%%%%%%%%%%%%%%%%%%%%%%%%%%%%%%%%%%%%%%%%%%%%%%%%%%%%%%%%%%%%%%%%%%%%%%%%%%%
\section{Discussion}

In this article, we deal with COVID-19 data to study the trend of the epidemic at the global situation. We Choose the mortality rate as an appropriate metric which measures the relative relation between the cumulative confirmed cases and death cases. The significant growth in the mortality rate in a short period motivates us to detect whether the trend of mortality rate changes. From the time-dynamic of density of mortality rate, we find that the mode of trend seems to have an obvious change at 78 days after first day when the reported death cases exceed 30. Then we discuss the reasons of abrupt change of mortality rate trend from aspects both subjective and objective. The discussion indicates that various factors impose impacts on the epidemic situation, including the policies and measures each country and region adopted for prevention before and after, with the social impacts caused by them, the medical conditions and services in countries with different levels of development, and more importantly, the mutation of virus strain itself.

Further analysis work can be conducted on the COVID-19 data. Since from the discussion of the global corona virus situation, some countries occupy important positions to affect the change of overall trend. Though many governments have made great efforts to control the pandemic in their countries, the rapid deterioration of corona virus in some North American and European countries still made the global situation less optimistic. Therefore, it is natural to make further investigations including the epidemic trends in these representative countries and regions with their potential factors, and comparison with the global situation, or other countries with better control of situations.
%%%%%%%%%%%%%%%%%%%%%%%%%%%%%%%%%%%%%%%%%%%%%%%%%%%%%%%%%%%%%%%%%%%%%%%%%%%%%%%%%%%%%%%%%%%%%%%%%%%%%%%%%%%%%%%%%%
%%%%%%%%%%%%%%%%%%%%%%%%%%%%%%%%%%%%%%%%%%%%%%%%%%%%%%%%%%%%%%%%%%%%%%%%%%%%%%%%%%%%%%%%%%%%%%%%%%%%%%%%%%%%%%%%%%
%\bibliographystyle{apalike}
%\bibliography{cpbib}

\begin{thebibliography}{}

\bibitem[Arlot et~al., 2012]{ach12}
Arlot, S., Celisse, A., and Harchaoui, Z. (2012).
\newblock Kernel change-point detection.
\newblock {\em arXiv preprint. arXiv:1202.38786.}

\bibitem[Carroll et~al., 2020]{cbc20}
Carroll, C., Bhattacharjee, B., and Chen, Y.~e. (2020).
\newblock Time dynamics of covid-19.
\newblock {\em medRxiv}.

\bibitem[Chen and Friedman, 2017]{cf17}
Chen, H. and Friedman, J.~H. (2017).
\newblock A new graph-based two-sample test for multivariate and object data.
\newblock {\em Journal of the American Statistical Association}, 112:397--409.

\bibitem[Chen and Zhang, 2015]{cz15}
Chen, H. and Zhang, N. (2015).
\newblock Graph-basd change-point detection.
\newblock {\em The Annals of Statistics}, 43:139--176.

\bibitem[Chu and Chen, 2019]{cc19}
Chu, L. and Chen, H. (2019).
\newblock Asymptotic distribution-free change-point detection for multivariate
  and non-euclidean data.
\newblock {\em The Annals of Statistics}, 47:382--414.

\bibitem[De~Ridder et~al., 2016]{dvr16}
De~Ridder, S., Vandermarliere, B., and Rycebusch, J. (2016).
\newblock Detection and localization of change points in temporal networks with
  the aid of stochastic block models.
\newblock {\em Journal of Statistical Mechanics: Theory and Experiment},
  113302.

\bibitem[Desobry et~al., 2005]{ddd05}
Desobry, F., Davy, M., and Doncarli, C. (2005).
\newblock An online kernel change detection algorithm.
\newblock {\em IEEE Transactions on Signal Processing}, 53(8):2961--2974.

\bibitem[Dubey and Muller, 2020]{dm20}
Dubey, P. and Muller, H.~G. (2020).
\newblock Fr\'echet change-point detection.
\newblock {\em The Annals of Statisics(to appear)}.

\bibitem[Garreau and Arlot, 2018]{ga18}
Garreau, D. and Arlot, S. (2018).
\newblock Consistent change-point detection with kernels.
\newblock {\em Electroic Journal of Statistics}, 12:4440--4486.

\bibitem[Hirose, 2020]{h20}
Hirose, H. (2020).
\newblock A relationship between sir model and generalized logistic
  distribution with applications to sars and covid-19.
\newblock {\em arXiv preprint. arXiv:2009.09653}.

\bibitem[James and Menzies, 2020]{jm20}
James, N. and Menzies, M. (2020).
\newblock Covid-19 in the united states:trajectories and second surge behavior.
\newblock {\em arXiv preprint. arXiv:2008.02068}.

\bibitem[Kawahara and Sugiyama, 2012]{ks12}
Kawahara, Y. and Sugiyama, M. (2012).
\newblock Sequential change-point detection based on direct density-ratio
  estimation.
\newblock {\em Statistical Analysis and Data Mining}, 5(2):114--127.

\bibitem[Kawahara et~al., 2007]{kym07}
Kawahara, Y., Yairi, T., and Machida, k. (2007).
\newblock Change-point detection in time-series data based on subspace
  identification.
\newblock {\em In: Proceedings of the 7th IEEE International Conference on Data
  Mining}, 559-564.

\bibitem[Mathur and Shaw, 2020]{ms20}
Mathur, N. and Shaw, G. (2020).
\newblock An empirical model on the dynamics of covid-19 spread in human
  population.
\newblock {\em arXiv preprint. arXiv:2008.06346}.

\bibitem[Moskvina and Zhigljavsky, 2003]{mz03}
Moskvina, V. and Zhigljavsky, A. (2003).
\newblock Change-point detection alogrithm based on sigular-spectrum analysis.
\newblock {\em Communications in Statistics: Simulation and Computation},
  32:319--352.

\bibitem[Peel and Clauset, 2015]{pc15}
Peel, L. and Clauset, A. (2015).
\newblock Detecting change points in the large scale structure of evolving
  networks.
\newblock {\em In AAAI}, 15:1--11.

\bibitem[Petersen and Muller, 2016]{pm16}
Petersen, A. and Muller, H.~G. (2016).
\newblock Functional data analysis for density functions by transformation to a
  hilbert space.
\newblock {\em The Annals of Statisics}, 44:183--218.

\bibitem[Petersen and Muller, 2019]{pm19}
Petersen, A. and Muller, H.~G. (2019).
\newblock Wasserstein covariance for multiple random densities.
\newblock {\em Biometrika}, 106:339--351.

\bibitem[Rayskin, 2020]{r20}
Rayskin, V. (2020).
\newblock Multivariate time series approximation by multiple trajectories od a
  dynamical system. applications to internet traffic and covid-19 data.
\newblock {\em arXiv preprint. arXiv:2002.05326}.

\bibitem[Rodrigues and Helene, 2020]{rh20}
Rodrigues, T. and Helene, O. (2020).
\newblock A monte carlo approach to model covid-19 deaths and infections using
  gompertz functions.
\newblock {\em arXiv preprint. arXiv:2008.04989}.

\bibitem[Roy and Karmakar, 2020]{rk20}
Roy, A. and Karmakar, S. (2020).
\newblock Time-varying auto-regressive models for count time series.
\newblock {\em arXiv preprint. arXiv:2009.07634}.

\bibitem[Takeuchi and Yamanishi, 2006]{ty06}
Takeuchi, Y. and Yamanishi, K. (2006).
\newblock A unifying framework for detecting outliers and change points from
  non-stationary time series data.
\newblock {\em IEEE Transactions on Knowledge and Data Engineering},
  18(4):482--489.

\bibitem[Tang et~al., 2020]{twz20}
Tang, C., Wang, T., and Zhang, P. (2020).
\newblock Functional data analysis : an application to covid-19 data in the
  united states.
\newblock {\em arXiv preprint. arXiv:2009.08363}.

\bibitem[Wang et~al., 2018]{wyr18}
Wang, D., Yu, Y., and Rinaldo, A. (2018).
\newblock Optimal change point detection and localization in sparse dynamic
  networks.
\newblock {\em arXiv preprint arXiv:1809.09602}.

\bibitem[Zhang and Lin, 2020]{zl20}
Zhang, T. and Lin, G. (2020).
\newblock Generalized k-means in glms with applications to the outbreak of
  covid-19 in the united states.
\newblock {\em arXiv preprint. arXiv:2008.03838}.

\end{thebibliography}

\end{document}